\documentstyle[12pt,aasms4]{article}

\begin{document}

\def\msun {M$_{\odot}$~}  
\def\es{elliptical galaxies{ }}
\def\etal{{\it et al.}{ }}
\def\mgs{Mg$_b-\sigma${ }}
\def\mg{Mg$_b$ line strength{ }}
\def\veldis{velocity dispersion{ }}
\def\lum{luminosity{ }}

\title{Exploring Cluster Ellipticals as Cosmological Standard Rods\altaffilmark{1,2} }

\author{Ralf Bender\altaffilmark{3}, R.P. Saglia, Bodo
Ziegler\altaffilmark{3}, Paola Belloni, Laura Greggio\altaffilmark{4},
Ulrich Hopp}
\affil{Universit\"ats-Sternwarte, Scheinerstra\ss e 1,
        81679~M\"unchen, Germany  }

\author{Gustavo Bruzual}
\affil{ CIDA, Apartado Postal 264, M\'erida 5101-A, Venezuela }

\altaffiltext{1}{Based on observations with the NASA/ESA {\it Hubble
Space Telescope}, obtained at the Space Telescope Science Institute,
which is operated by AURA, Inc., under NASA contract NAS 5-26555.}

\altaffiltext{2}{Based on observations carried out at the European
Southern Observatory, La Silla, Chile. }

\altaffiltext{3}{Visiting Astronomer of the German-Spanish
Astronomical Center, Calar Alto, operated by the Max-Planck-Institut
f\"ur Astronomie, Heidelberg, jointly with the Spanish National
Commission for Astronomy}

\altaffiltext{4}{On leave from Dipartimento di Astronomia,
Universit\`a di Bologna, I-40100 Bologna, Italy}

\begin{abstract}

We explore the possibility to calibrate massive cluster ellipticals as
cosmological standard rods using the Fundamental Plane relation
combined with a correction for luminosity evolution. Though cluster
ellipticals certainly formed in a complex way, their passive evolution
out to redshifts of about 1 indicates that basically all major merging
and accretion events took place at higher redshifts. Therefore, a
calibration of their luminosity evolution can be attempted. We propose
to use the Mg$-\sigma$ relation for that purpose because it is
independent of distance and cosmology. We discuss a variety of
possible caveats, ranging from dynamical evolution to uncertainties in
stellar population models and evolution corrections to the presence of
age spread. Sources of major random and systematic errors are analysed
as well. 

We apply the described procedure to nine elliptical galaxies in two
clusters at $z=0.375$ and derive constraints on the cosmological
model.  For the best fitting $\Lambda$-free cosmological model we
obtain: $q_o \approx 0.1$, with 90\% confidence limits being $0 < q_o
< 0.7$ (the lower limit being due to the presence of matter in the
Universe). If the inflationary scenario applies (i.e. the Universe has
flat geometry), then, for the best fitting model, matter and $\Lambda$
contribute about equally to the critical cosmic density
(i.e. $\Omega_m \approx \Omega_\Lambda \approx 0.5$). With 90\%
confidence $\Omega_\Lambda$ should be smaller than 0.9.

\end{abstract}


\section{Introduction}

A central issue of modern cosmology is the determination of the
geometry of the Universe, or, equivalently of its density parameter
$\Omega$ and of the cosmological constant $\Lambda$.  In recent years,
several lines of argument indicate that $\Omega$ may be smaller than
1, and, therefore, the universe may be open and have negative
curvature. On the other hand, if the inflationary scenario applies
(see, e.g., \cite{Linde90}),
then the Universe should have flat geometry, i.e., either the matter
density has the critical value ($\Omega_m=1$), or, if $\Omega_m<1$,
$\Omega_m+\Omega_\Lambda=1$ with $\Omega_\Lambda=\Lambda/(3H_o^2)$.
In the latter case, a non-zero cosmological constant is needed which
may have interesting consequences for the cosmological deceleration
parameter $q_o=\Omega_m/2-\Omega_\Lambda$, i.e., $q_o$ may be negative.

Most methods to determine the deceleration parameter $q_o$ directly
rely on the variation of the {\it apparent} sizes or brightnesses of
standard objects with redshift (\cite{Sanda95}). So far, all
measurements were inconclusive, because all objects which were bright
or large enough to be potential standard rods or candles
(e.g. galaxies or clusters of galaxies), show significant evolution
with redshift. In fact, until now most attempts to measure $q_o$ were
better tests for galaxy and cluster evolution than for the
cosmological model. Only recently, Supernovae Ia, which may be true
standard candles, have raised hopes for tighter constraints on $q_o$
(e.g., \cite{GP95,Leib96}). The ongoing SNIa programs promise indeed
to provide significant constraints on $q_o$ (e.g., \cite{Perl97}), once
enough SNIa will have been found at intermediate redshifts and all
observational caveats will have been understood.

In this paper we want to explore whether luminous cluster ellipticals
can be calibrated as cosmological standard rods.  Though it is almost
certain that luminous elliptical galaxies have experienced a complex
formation history (with events ranging from accretion of satellite
galaxies to violent similar--mass mergers, see
e.g. \cite{Bende90,Schwe90}), most of these events took place early
in their evolution. This follows both from the homogenous properties
of local cluster ellipticals (\cite{D87,BLE92,BBF93,RC93}) and from the
very small redshift evolution of luminosities (\cite{GPMC95,LTHCF95}),
colors (\cite{SED97,Eetal97}), and spectral indices (\cite{BZB96}).
Furthermore, the redshift evolution of the Fundamental Plane relation
of ellipticals (\cite{DD87,FDDBLTW87b,BBF92}) has also been found to
be consistent with passive evolution of the stellar populations in
ellipticals (\cite{DF96,KDFIF97}). Basically passive evolution of
cluster ellipticals is even expected in the most rapidly evolving
standard cold dark matter model, where neither significant star
formation nor accretion processes take place at $z<0.5$
(\cite{KC97}). The star formation activity observed in distant
clusters of galaxies seems to be confined to blue infalling and/or 
harrassed disk galaxies (\cite{ODB97,MLK97,Bal97}) 
which may turn into intermediate
luminosity S0s today but are unlikely to alter the population of
massive cluster ellipticals at a significant level.

Because massive cluster ellipticals evolve only passively up to
redshifts $z=0.5...1$, one can attempt to calibrate them as
cosmological standard rods.  The method proposed here is based on the
Fundamental Plane relation of elliptical galaxies, which allows us to
derive accurate radii from their velocity dispersions and surface
brightnesses, combined with a correction for luminosity evolution
obtained from the Mg$-\sigma$ test (\cite{BZB96}).  We give a first
application of this method by deriving constraints on $q_o$ from nine
ellipticals in two clusters at $z=0.375$. In \S 2 we present the
method, in \S 3 we discuss possible caveats. In \S 4 observations and
data analysis are described, the results of which are presented in \S
5.  Conclusions are drawn in \S 6.

\section{Calibrating elliptical galaxies as standard rods}

The FP describes the observed tight scaling relation between effective
radius ($R_e$), mean effective surface brightness ($\langle$SB$\rangle_e$) 
and velocity dispersion ($\sigma$) of cluster ellipticals:
\begin{equation}
\log R_e = 1.25\log\sigma + 0.32\langle{\rm SB}\rangle_e - 8.895
\end{equation}
with $R_e$ in kpc, $\sigma$ in km/s, $\langle$SB$\rangle_e$ in the
B-band. The constant -8.895 was derived from the Coma cluster
ellipticals with a Hubble constant of H$_o = 50$km/s/Mpc, the error in
the constant is about 0.01 (see below). The FP allows to predict the
effective radii $R_e$ of elliptical galaxies with better than 15\%
accuracy from their velocity dispersions and surface brightnesses,
i.e., a distance dependent quantity ($R_e$) can be estimated from two
distance independent quantities ($\sigma$, $\langle$SB$\rangle_e$).
If a large enough number of ellipticals are measured per cluster, then
the distance to the corresponding cluster can be determined with
significantly better than 5\% accuracy because otherwise the observed
peculiar velocities of clusters would be much larger (e.g.
\cite{Burst89,JFK96}).

As argued in the introduction, we have strong evidence that massive
ellipticals in rich clusters evolve only passively between now and at
least $z\approx 0.5$.  Passive evolution of the stellar population can
be accurately measured with the Mg$-\sigma$ relation which is
independent from $q_o$ and H$_o$ (\cite{BZB96}), and is within current
limits the same for clusters of similar richness/velocity dispersion
(\cite{Joerg97}, see also below). In local cluster ellipticals, the
Mg$_b$ absorption is tightly coupled to the velocity dispersion
$\sigma$ of the galaxy constraining the scatter in age and metallicity
at a given $\sigma$ to be smaller than 15\%.  Therefore, in the case
of passive evolution, Mg$_b$ decreases with redshift only because the age
of the population decreases.  One can use population synthesis models
to translate the observed ${\rm Mg}_b$ weakening into an estimate of
the B--band luminosity evolution.  Based on models by
\cite{Worth94,BC97}, and \cite{BCT96}, we obtain consistently $\Delta B = 1.4
\Delta$Mg$_b$ for a Salpeter initial mass function, metallicities
between 1/3 and 3 times solar and ages between 3~Gyr and 15~Gyr (see
Figure~1). The slope of 1.4 shows no dependence (within 0.1) on
evolutionary tracks and other differences in the synthesis models,
which demonstrates that this differential comparison of Mg$_b$ and
B-band evolution is quite robust (see also \S 3). 

\placefigure{fig1}

\section{Caveats}

Obviously, there is a number of caveats that have to be
addressed at this point. Most of them need further checking 
before ellipticals can reliably serve as standard rods.
However, the influence of all these effects is rather small
and there is justified hope that they may be controlled better
in the future. 

(1) Dynamical evolution. We now have very good evidence that dynamical
evolution will not alter the FP at a detectable level. There are two simple
observational arguments. First, different types of objects, like anisotropic
ellipticals, rotationally flattened ellipticals and S0 galaxies show no
significant offsets from the mean FP relation. This means that objects of
very different internal structure and dynamics and very different
formation/evolution histories are indistinguishable in the edge--on view of
the FP, implying that changes in the dynamical structure transforming
objects from one type to an other will not affect the FP relation. Second,
the successful use of the FP to derive peculiar motions from clusters and
groups of galaxies of different richness and internal evolution
(\cite{BFD90}) indicates that environment and interaction can alter the
fundamental plane parameters on the few percent level at most. Finally,
recent numerical simulations show that dissipationless merging of objects
within the FP produces objects that are again in the plane (\cite{CCC95}).

(2) Dependence on environment, i.e., systematic differences between the
fundamental plane and Mg$-\sigma$ relations of different clusters. At
present, the small observed peculiar velocities of clusters (e.g.
\cite{Burst89,JFK96}) imply that the systematic variation of the FP zeropoint
from cluster to cluster must be significantly smaller than 5\%. For the
Mg$-\sigma$ relation a weak dependence on cluster velocity dispersion or
richness has been observed (\cite{Joerg97}, but see Colless et al., in
preparation).  In any case, there is no evidence
that, at a given richness, the zeropoint of the Mg$_b-\sigma$ relation varies
by more than 0.08\,\AA. Furthermore, the effects of different zeropoints can
be minimized by observing a large enough sample of ellipticals in different
clusters. 

(3) Sample selection effects. We estimated the combined biasing effect
of the sample size, sample selection and allowed range of the FP
parameters by means of Monte Carlo simulations following
\cite{Sagli97}. We find that the systematic bias of the FP zero-point
is smaller than the random variation estimated in Table 1, for the
range of FP coefficients given by J\o rgensen et al. (1996). A large
complete sample of distant cluster ellipticals can minimize 
selection effects.

(4) Dust absorption. Dust absorption can be estimated by comparing
colors and line-strength indices.  We could rule out a significant
presence of dust in the $z=0.375$ ellipticals by checking the
relation between their rest--frame B$-$V color and their Mg$_b$
values, finding that it is consistent with the local (B$-$V)--Mg$_b$
relation (e.g., \cite{BBF93}) and population synthesis models
(\cite{Worth94}). 

(5) Population synthesis models. At present, population synthesis
models (e.g., \cite{Worth94,BC97,BCT96}) do not always give a
consistent interpretation of ages and metallicities of stellar
populations. Close inspection shows that the differences mostly arise
because of different zeropoints in the modeled magnitudes, colors and
line-strengths for a given age-metallicity combination.  However, the
different models agree very well with respect to {\it differential}
relations. E.g., dMg$_b$/dB as a function of age is virtually the same for all models,
despite of different stellar evolution tracks and numerical codes (see
Figure~1). At present, the biggest uncertainty with respect to the
method discussed here stems from the fact that all synthesis codes
rely on the same set of fitting-functions for the line-indices, introduced by
\cite{GFBGCP93} and revised by \cite{Worth94}. These reproduce the
Mg-indices of G- and K-stars as a function of temperature, gravity and
metallicity in an excellent manner, but only for stars with solar
abundance ratios while massive ellipticals are overabundant in Mg
relative to Fe.  It is plausible that the {\it differential}
temperature dependence of the Mg$_b$ index is not very sensitive to
this effect but further checking is needed.

(6) Initial mass function.  The relation between time evolution in the
B-band and Mg$_b$ is dependent on the exponent $x$ of the initial mass
function.  From models of \cite{BC97} we find: $\Delta B = 1.4
(1-0.3(x-x_s)) \Delta{\rm Mg}_b$, where the subscript $s$ indicates
the Salpeter exponent ($x_s=1.35$).  At present, there is neither
evidence nor a convincing physical argument for an IMF slope much
different from Salpeter's for the metallicities around solar and for
the mass range concerned (between 0.8$M_\odot$ and 1.2$M_\odot$).
Still, this is a key assumption for the method to work. 

(7) Effects of mixed populations. We performed tests with varying
fractions of intermediate age populations superposed on an old
population, all constrained in a way that the objects at $z=0.375$
would {\it not} be regarded as of $E+A$ type from spectral
characteristics (\cite{DG83}).  For plausible population mixes, the
uncertainty in the evolution correction is smaller than 0.1~mag in the
B--band. Figure~1 shows two examples for the Mg$_b$ vs. B-band evolution of
mixed populations. One experienced a 10\% starburst 3~Gyrs, the other
1.5~Gyrs before the redshift of observation ($z=0.4$). In the second
case a significant curvature is present in the Mg$_b$-B-relation.
This indicates that evolution effects due to mixed populations with
$\Delta B < -0.6$~mag are more difficult to correct reliably. However,
note that in case of massive ellipticals it is very difficult to add
10\% of young stars over a short period because the gas content of
most objects that can be accreted is too small relative to the total
mass of the elliptical.

\section{A First Application: Observations and Data Analysis}

We analysed a sample of nine elliptical galaxies in the clusters
Abell~370 and MS~1512+36 at $z=0.375$. Abell~370 is of similar
richness as the Coma Cluster, which serves as our local calibrator, 
MS~1512+36 is less rich. 

Spectroscopic data were obtained at the Calar Alto 3.5m telescope and
the NTT at ESO. Details of the spectroscopic observations and data
analysis can be found in Ziegler \& Bender (1997). The largest uncertainty
in the obtained velocity dispersions and Mg$_b$-indices comes from
the applied aperture corrections necessary to compare local and
distant ellipticals. However, the errors from aperture corrections can
be expected to become smaller in the near future because better data
for nearby ellipticals will allow to integrate over apertures large
enough for direct comparison with distant ellipticals, and spatial
resolution will also improve for distant ellipticals due to the use of
large telescopes. The spectroscopic parameters for the local
comparison sample were taken from \cite{FWBDDLT89}. At present, there
still exists some uncertainty in the parameters of the low-redshift
Mg$_b-\sigma$ relation (\cite{D87}, see Figure~2) because the Mg$_2$
values for Coma and Virgo ellipticals have to be tansformed to Mg$_b$
values. However, this uncertainty is not crucial at the present stage
of analysis and will be reduced by future measurements. We used the
relation Mg$_b = 15.0$Mg$_2$, see Ziegler \& Bender (1997).

Accurate radii and surface brightnesses for the distant ellipticals
were obtained with the Hubble Space Telescope (HST) and the
refurbished Wide Field and Planetary Camera (WFPCII).  The structural
parameters $R_e$ and $\langle{\rm SB}\rangle_e$ in the F675W filter
were derived using the two--component fitting algorithm developed by
\cite{SBBBCDMW97}, with HST PSF convolution tables. Ground--based
imaging at the ESO NTT and the Calar Alto 2.2m telescope in several
colors (Ziegler, in preparation) complemented the HST imaging and
allowed an accurate transformation to rest--frame colors using models
of \cite{BC97}. Surface brightnesses were transformed to rest--frame
B--band, corrected for cosmological dimming ($(1+z)^4$) and for
passive evolution using the ${\rm Mg}_b-\sigma$ relation described
above.  Galactic extinction was corrected using extinctions of
\cite{BH84}, kindly provided by David Burstein.

The photometric parameters of the local comparison sample of Coma
galaxies (\cite{SBD93}) were re--derived with the same procedure as
used for the $z=0.375$ objects. The B--band photometric zeropoints
were improved using aperture photometry from \cite{LVC83} and
\cite{JFK95} reducing the rms scatter about the Coma FP to about 10\%
in the effective radius (see Figure~3).  We applied K--corrections and
corrections for galactic extinction as in \cite{FWBDDLT89} plus the
correction for the cosmological dimming of surface brightness.  Since
the $z=0.375$ ellipticals are about a factor of 10 more distant than
the Coma galaxies, the factor 10 better sampling and smaller PSF of
HST relative to the ground--based imaging of Coma corresponds to a
perfectly matched instrumental set--up.

{\bf Error analysis.} An account of the major sources of error is
given in Table 1. The dominant error is due to the correction for
luminosity evolution. This error is partly random and partly
systematic. For the rather few objects we consider here, the random
error is still dominating. Note that because of error coupling
in the Kormendy relation $R_e \propto  I_e^{0.8}$ ($I_e =$ surface
brightness in linear flux units), the error in  $R_e I_e^{-0.8}$
is much smaller than in each of $R_e$ and $I_e$ separately
(\cite{Kor77,Kor87}).

The fact that the scatter of the $z=0.375$
ellipticals around the FP is somewhat smaller (only 15\%) than we
expect from the error estimate ($\approx 20$\%) suggests that our
error estimate is conservative.

\placetable{tbl-1}

\placetable{tbl-2}

\placefigure{fig2}

\placefigure{fig3}

\section{Results}

The photometric and spectroscopic parameters for the elliptical
galaxies in the clusters Abell~370 and MS~1512+36 at $z=0.375$ are
given in Table 2.  Figure~2 shows the Mg$_b-\sigma$ relation for the
local comparison sample of Coma and Virgo ellipticals and the nine
ellipticals observed in Abell~370 and MS1512+36 at $z=0.375$. The
$z=0.375$ objects show a Mg$_b$ weakening of about 0.345\AA\ which is
typical for ellipticals at this redshift (\cite{BZB96}).

Figure~3 shows the edge--on view of the Fundamental Plane for Coma
ellipticals with $\sigma >120$km/s and for the $z=0.375$ ellipticals.
The angular distances at which a perfect match between the two samples
is achieved are 139~Mpc/h$_{50}$ for Coma and 1400~Mpc/h$_{50}$ for
the $z=0.375$ ellipticals (h$_{50}$ is the Hubble constant in units of
50km/s/Mpc), see below.  The $z=0.375$ objects are shown three times
in Figure~3, first as observed in the rest-frame B-band, then with a
correction applied for cosmological $(1+z)^4$ dimming, and finally
corrected for luminosity evolution. The luminosity correction is
calculated based on the {\it mean} offset in the Mg$_b$ absorption of
the local and the $z=0.375$ sample, i.e., $\Delta \langle{\rm
SB}\rangle_{B,e} = 1.4 \Delta{\rm Mg}_b = 0.48$mag/arcsec$^2$. We
could also have corrected the luminosity evolution of the objects
individually and would have obtained the same result. In fact, the
residuals of the distant ellipticals from the fundamental plane and
from the Mg$_b-\sigma$ relation correlate with each other in the
expected way, though the error bars are large (see Figure~4).  
This supports the idea
that the evolution we see is really due to age. While for local
samples of ellipticals we cannot conclude reliably that the residuals
from the FP and Mg$_b-\sigma$ relations are correlated and caused by
age (see \cite{JFK96}), the effects of age spread must increase with
redshift and may lead to the correlation observed in Figure~4.

\placefigure{fig4}

The cosmological model is constrained on the basis of Figure~3 as
follows. The distance ratio at which the fully corrected FP of the
$z=0.375$ clusters matches the FP of Coma is $10.1 \pm 0.8$. With
an angular distance of 139~Mpc/h$_{50}$ for Coma (corrected to the
Cosmic Microwave Background rest--frame following \cite{FWBDDLT89})
this ratio corresponds to an angular distance of 1400~Mpc/h$_{50}$ for
Abell~370 and MS1512+36 at $z=0.375$ which is the distance adopted in
Figure~3 (h$_{50}$ is the Hubble constant in units of
50~km/s/Mpc). For a plausible range of cosmological models, the
distance to the Coma cluster is, because of its proximity, virtually
independent from $q_o$ (at the level of 0.5\%), while the distance to
the $z=0.375$ clusters varies by more than 10\%.  Since the geometry
of the Universe is determined by the ratio of distances, $q_o$ is
independent from the Hubble constant.  The relative error of mean
distance to the $z=0.375$ clusters is about 8\% (see Table 1). The
mean distance to $z=0.375$ and its error give immediate constraints on
the cosmological model, see Figure~5. If the cosmological constant
vanishes then our measurements, together with the observational fact
that there exists matter in the Universe, constrain the cosmological
density parameter to be $0 < \Omega_m < 1.4$, or the cosmological
deceleration parameter to be $0 < q_o < 0.7$, with 90\% confidence. If
the Universe has flat geometry as suggested by inflation, then the
preferred model would have $\Omega_m = \Omega_\Lambda = 0.5$.
$\Omega_\Lambda$ is constrained to fall in the range $-0.25 <
\Omega_\Lambda < +0.9$, again with 90\% confidence.

\placefigure{fig5}

\section{Conclusions}

We have explored the possibility to
calibrate elliptical galaxies as cosmological standard rods.  The
proposed method is based on the fundamental plane relation of elliptical
galaxies and the Mg$_b-\sigma$ test which allows us to calibrate the
luminosity evolution of their stellar populations. The main
assumptions that have to be made for this procedure to work are: (1)
the fundamental plane and Mg$-\sigma$ relations are the same for
clusters of similar richness and (2) the slope of the initial mass
function of low mass stars is known and has the same value in
different objects and environments (here we adopt the Salpeter value).
Further critical issues that need future checking are the reliability
of stellar population models and the influence of sample selection
effects and mixed populations.

A first application of the method has been given. We have shown that,
under the provisos given above, it is possible to derive interesting
constraints on the matter density and the cosmological constant using
elliptical galaxies.  In the future, stronger constraints should be
possible if large samples of ellipticals up to redshift of close to 1
(\cite{KDFIF97}) are analysed in a similar way, and if the caveats and
systematic effects discussed above can be controlled efficiently.

\acknowledgements
Thanks go to Dr. R. Carlberg who provided redshifts for 
members of MS~1512$+$36 and to Dr. D. Burstein who calculated
galactic extinctions for Abell~370 and MS~1512$+$36. We thank the
anonymous referee for helpful comments. This work was
funded by the Sonderforschungsbereich 375 of the DFG and by DARA grant
50 OR 9608 5.  RB acknowledges additional support by the
Max--Planck--Gesellschaft.  LG was partially supported by the
Alexander-von-Humboldt foundation, GB by the EU under CiL-CT93-0328VE.

\clearpage


\clearpage

\figcaption[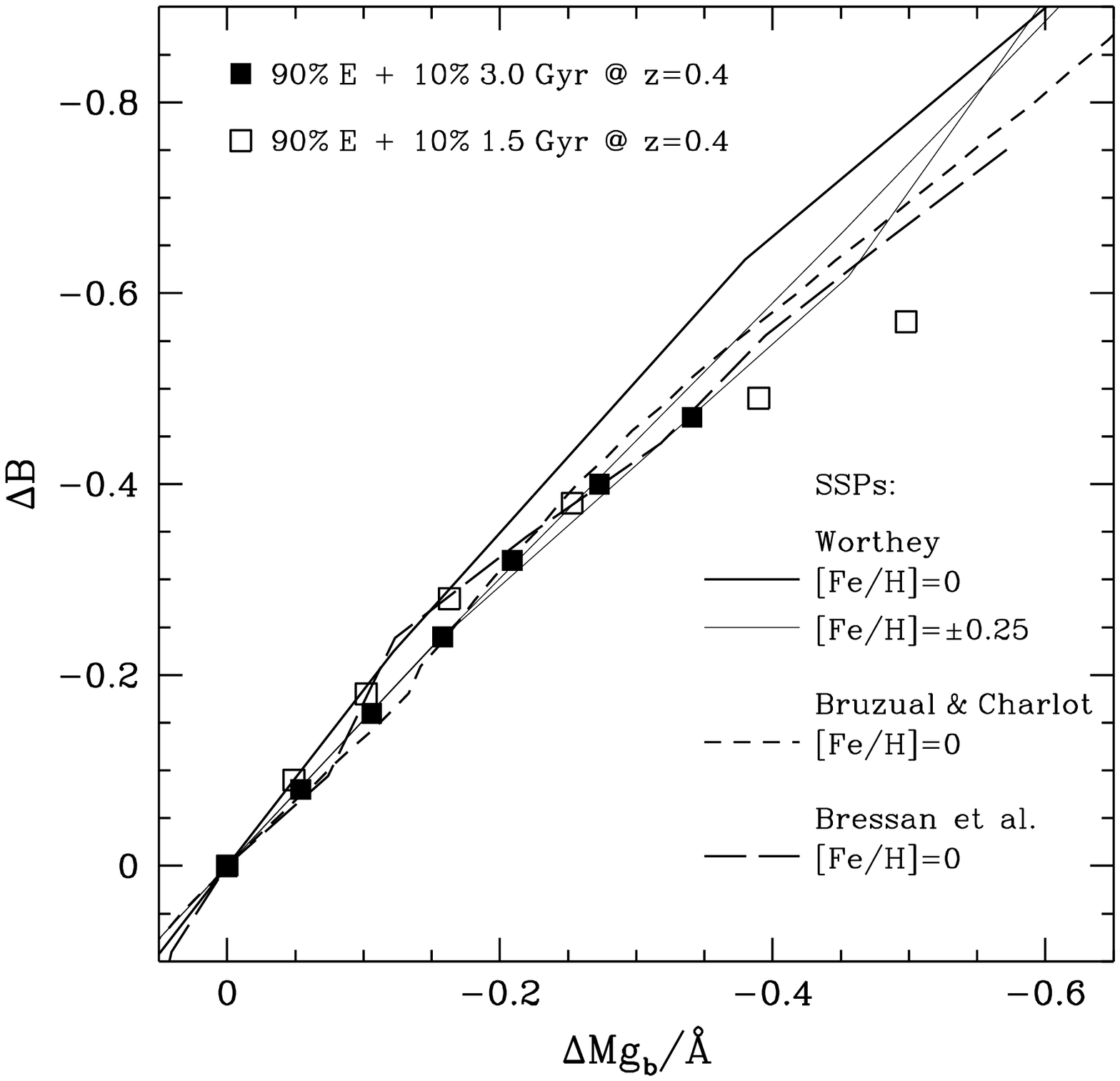]{Differential relation between Mg$_b$
absorption and B-band luminosity as a function of age for different
metallicities.  Lines for different metallicities were normalized to the
same B-band luminosity at an age of 12~Gyr and refer to simple stellar
populations. Composite stellar populations are indicated via
open and filled squares. The open (filled) squares trace the
differential evolution of a single galaxy that consists to 90\% of an
old population (E) to which a starburst 1.5~Gyr (3~Gyr) before
observation added a mass of 10\%. \label{fig1} }

\figcaption[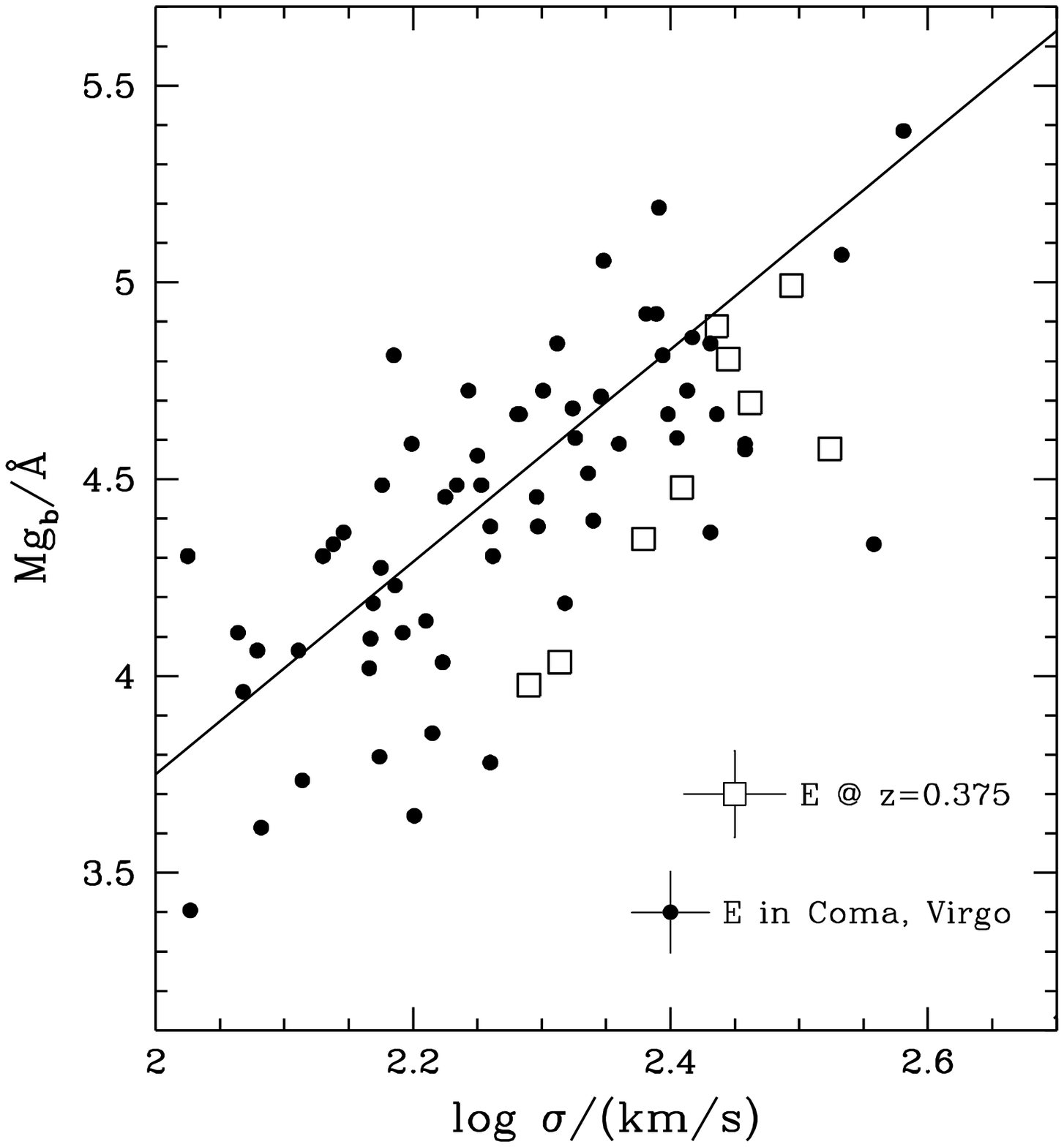]{The Mg$_b-\sigma$ relation for elliptical
galaxies in the Coma and Virgo clusters (filled circles) and for elliptical
galaxies in Abell~370 and MS1512+36, both at $z=0.375$ (open
squares). All measurements have been brought to consistent aperture sizes.
Typical error bars are given in the lower right. The best fit to the Coma
data is Mg$_b = 2.7\log\sigma -1.65$ with $\sigma$ in km/s. The lower
Mg$_b$ of the distant ellipticals at a given $\sigma$ is due to
the (passive) evolution of  their stellar populations with redshift. \label{fig2} }

\figcaption[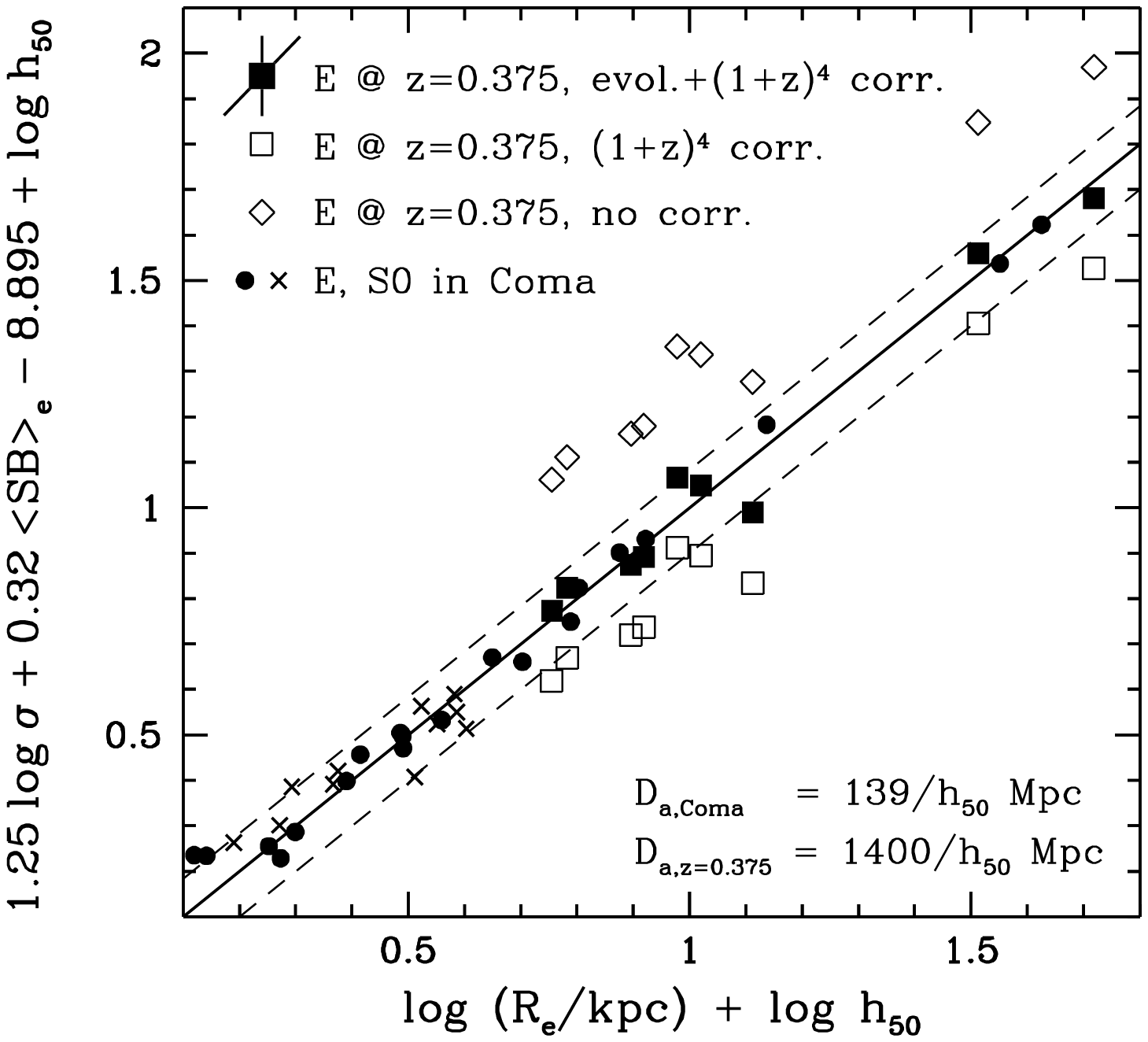]{The Fundamental Plane of
elliptical and S0 galaxies. Small filled circles and small crosses
denote elliptical galaxies and S0 galaxies in the Coma cluster with
velocity dispersions \hbox{$\sigma > 120~$km/s}, effective radii
$R_e$ in kpc and effective surface brightnesses $\langle$SB$\rangle_e$ in the
B band, h$_{50}$ is the Hubble constant in units of 50~km/s/Mpc.  The
typical measurement errors for the Coma objects are somewhat smaller
than the scatter indicates.  The large open diamonds show elliptical
galaxies in Abell~370 and MS1512+36, both at a redshift of $z =
0.375$, as observed and transformed to the B--band rest frame. The
large open squares show the $z = 0.375$ ellipticals after surface
brightness has been corrected for cosmological $(1+z)^4$ dimming. The
large filled squares represent the redshifted ellipticals after a
further correction for mean luminosity evolution has been applied. The
correction was derived from the offset in the Mg$-\sigma$ relations
between the distant and local ellipticals (see Figure 2).  A typical
error bar is shown for the $z = 0.375$ ellipticals in the upper
left. The FP relations match for an angular distance to Coma of
139~Mpc/h$_{50}$ and an angular distance of 1400~Mpc/h$_{50}$ to
Abell~370 and MS1512+36 at $z=0.375$.  The upper and lower dashed
lines show the mean FP relations of the $z=0.375$ objects if they were
at a distance of 1100~Mpc/h$_{50}$ or 1700~Mpc/h$_{50}$, respectively.
\label{fig3} }

\figcaption[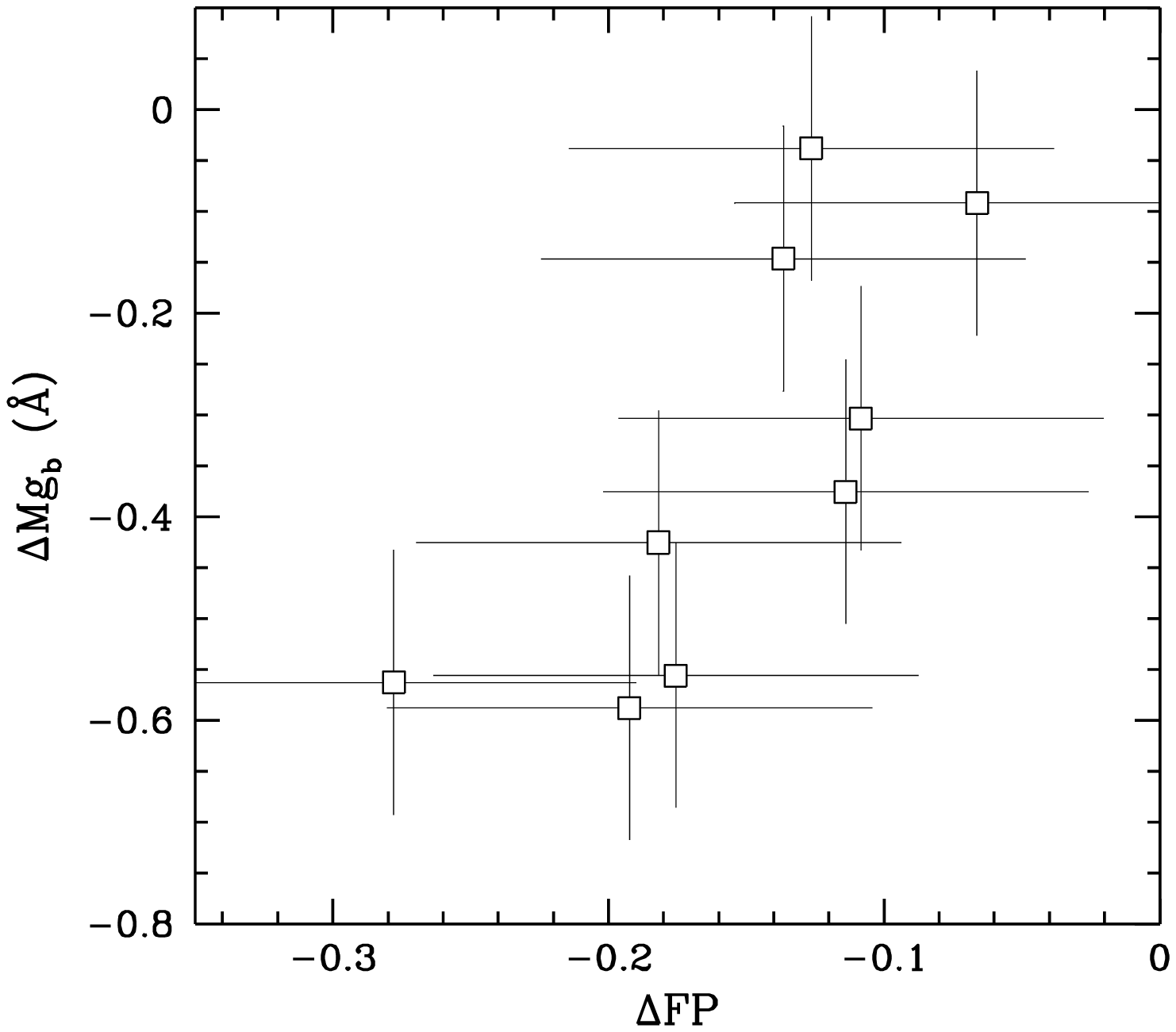]{The residuals from the local fundamental
plane {\it vs.} the residuals from the local Mg$_b-\sigma$ relation
for the $z=0.375$ ellipticals. Though the error bars are large, 
the residuals seem to correlate with each other. 
\label{fig4} }

\figcaption[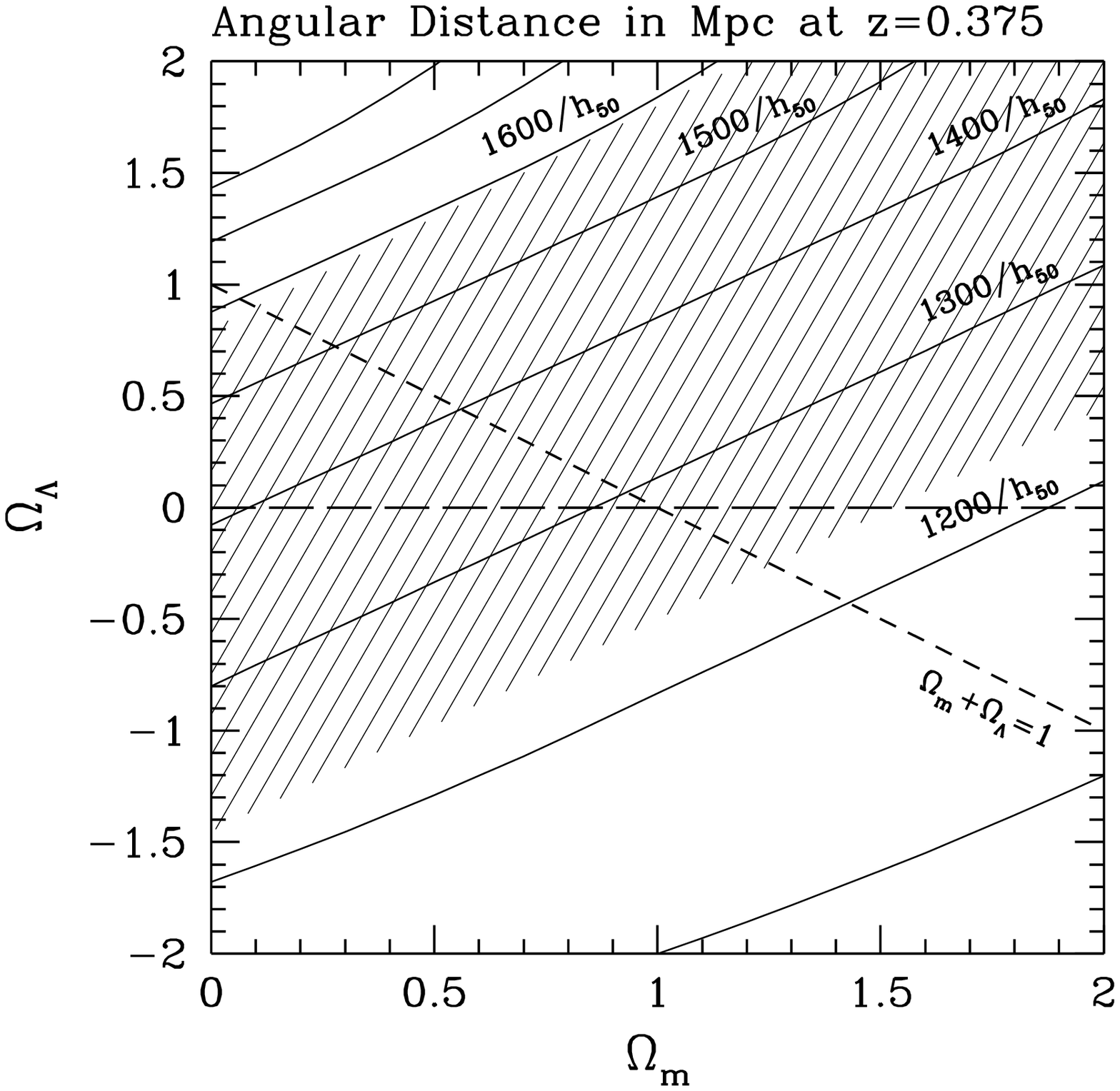]{The plane of matter density as measured by
$\Omega_m$ against $\Omega_\Lambda$ parametrizing the cosmological
constant.  Lines of constant angular distance for a redshift of 0.375
are shown.  The ellipticals observed at this redshift allow to
constrain the angular distance to $z=0.375$ which in turn defines a
probability strip in the $\Omega_\Lambda-\Omega_m$ plane. Values for
$\Omega_\Lambda$ and $\Omega_m$ which have $>90$\% likelihood lie in
the shaded area.  The horizontal long--dashed line corresponds to a
Universe without a cosmological constant, the diagonal short--dashed
line to a Universe with no curvature ($\Omega_m+\Omega_\Lambda=1$).
\label{fig5} }

\clearpage

\begin{table*}
\begin{center}
\begin{tabular}{lccc}
source of error~~~~~~~~~~~~~~~ & ~~~error for single object~~~  &
~~~error of mean for 9 objects     \\
\tableline
photometric ZP                & $\Delta{\rm SB}_e\;\; \approx 0.020$    &$\approx 0.020$    \\
transformation to rest--frame & $\Delta{\rm SB}_e\;\; \approx 0.030$    &$\approx 0.020$    \\
reddening correction          & $\Delta{\rm SB}_e\;\; \approx 0.020$    &$\approx 0.020$    \\
Kormendy product: $R_e I_e^{-0.8}$ & $\Delta{\rm SB}_e\;\; \approx 0.030$    &$\approx 0.011$    \\
evolution correction          & $\Delta{\rm SB}_e\;\; \approx 0.180$    &$\approx 0.064$    \\
velocity dispersion           & $\Delta\log\sigma \approx 0.040$        &$\approx 0.014$    \\
physical scatter in FP        & $\Delta{FP}\;\;\, \approx 0.040$        &$\approx 0.014$    \\
total error                   & $\Delta{FP}\;\;\, \approx 0.087$        &$\approx 0.032$    \\ 
\end{tabular}
\end{center}
\caption{Budget of major errors ($z=0.375$ clusters relative to Coma cluster),
with ${\rm FP} = 1.25 \log \sigma + 0.32 \langle{\rm SB}\rangle_e -
8.895$ and $\langle{\rm SB}\rangle_e = -2.5\log I_e + const.$
For a brief discussion of the error in the Kormendy product see text.
Note that the error of the mean is in some cases larger than the expected
statistical error because  of systematic effects.
\label{tbl-1}}
\end{table*}

\begin{table*}
\begin{center}
\begin{tabular}{lccccc}
object ~~~~ & ~~~R$_e$~~~ & ~~$\langle$SB$\rangle_e$~~ & ~~~log~R$_e$~~~ &
~~log~$\sigma$~~ & ~~~~Mg$_b$ ~~~~\\
 ~~~~ & arcsec & Bmag/arcsec$^2$ & kpc & km/s & \AA \\
\tableline
 Abell~370 \#13    &   0.830    &    20.18   &    0.755    &    2.445    &    4.805\\
 Abell~370 \#17    &   1.209    &    20.81   &    0.919    &    2.379    &    4.348\\
 Abell~370 \#18    &   0.884    &    20.48   &    0.783    &    2.409    &    4.479\\
 Abell~370 \#20    &   7.634    &    22.71   &    1.719    &    2.524    &    4.577\\
 Abell~370 \#23    &   1.387    &    20.90   &    0.978    &    2.494    &    4.992\\
 Abell~370 \#28    &   1.528    &    21.07   &    1.020    &    2.436    &    4.889\\
 Abell~370 \#32    &   1.889    &    21.37   &    1.112    &    2.314    &    4.035\\
 MS1512+36 \#09    &   4.755    &    22.57   &    1.513    &    2.462    &    4.694\\
 MS1512+36 \#11    &   1.147    &    21.10   &    0.896    &    2.290    &    3.977\\
\end{tabular}
\end{center}
\caption{Photometric and kinematic data for elliptical galaxies at
$z=0.375$. Object identification as in Ziegler \& Bender (1997).
Surface brightnesses are means within the effective radius, refer to
the rest-frame B-band and are corrected for extinction and
cosmological dimming, but not for evolution. Effective radii in kpc
were calculated with a distance of 1400~Mpc. Errors are given in
Table~1 and shown in the plots. \label{tbl-2}}

\end{table*}

\clearpage

\begin{figure}
\figurenum{Figure 1}
\plotone{fig1_final.eps}
\end{figure}

\clearpage

\begin{figure}
\figurenum{Figure 2}
\plotone{fig2_final.eps}
\end{figure}

\clearpage

\begin{figure}
\figurenum{Figure 3}
\plotone{fig3_final.eps}
\end{figure}

\clearpage

\begin{figure}
\figurenum{Figure 4}
\plotone{fig4_final.eps}
\end{figure}

\clearpage

\begin{figure}
\figurenum{Figure 5}
\plotone{fig5_final.eps}
\end{figure}

\end{document}